\documentstyle[psfig]{aa}

\def\hii{H{\sc ii}}

\def\sun{\hbox{$\odot$}}
\def\R23{\mbox{$\rm R_{23}$}}
\def\lya{\mbox{Ly-$\alpha$}}
\def\am{\mbox{\AA\,}}

\def\arcsec{\hbox{$^{\prime\prime}$}}

\def\kmsmpc{km s$^{-1}$ Mpc$^{-1}$}

\def\Hb{\mbox{${\rm H}{\beta}$}}
\def\Ha{\mbox{${\rm H}{\alpha}$}}
\def\OIII{\mbox{${\rm [O\,III]\,}{\lambda\lambda\,4959,5007}$}}
\def\OIIItemp{\mbox{${\rm [O\,III]\,}{\lambda\,4363}$}}
\def\OIIIa{\mbox{${\rm [O\,III]\,}{\lambda\,5007}$}}
\def\OIIIb{\mbox{${\rm [O\,III]\,}{\lambda\,4959}$}}
\def\OII{\mbox{${\rm [O\,II]\,}{\lambda\,3727}$}}

\def\NII{\mbox{${\rm [N\,II]\,}{\lambda\,6584}$}}
\def\NIIdob{\mbox{${\rm [N\,II]\,}{\lambda\lambda\,6548,6584}$}}

\begin{document}

\title{
The metallicity-luminosity relation at medium redshift based on faint
CADIS emission line galaxies\thanks{Based on observations
  obtained at the ESO VLT, Paranal, Chile; ESO programs 67.A-0175,
  68.B-0088, and 69.A-0266}}

\author{C. Maier\inst{1,2}, K. Meisenheimer\inst{1},
 H. Hippelein\inst{1}}

\offprints{C. Maier, \email{chmaier@phys.ethz.ch}}

\institute{
Max--Planck--Institut f\"ur Astronomie, K\"onigstuhl 17, 
           D-69117 Heidelberg, Germany
\and Department of Physics, Swiss Federal Institute of Technology (ETH Z\"urich), ETH H\"onggerberg, CH-8093, Z\"urich, Switzerland
}

\date{Received 3 December 2003 / Accepted  28 January 2004}

\abstract{The emission line survey within the \textbf{C}alar \textbf{A}lto \textbf{D}eep \textbf{I}maging
 \textbf{S}urvey (CADIS) detects galaxies with very low continuum brightness by using an imaging Fabry-Perot interferometer. With spectroscopic follow-up observations of $\rm{M}_{\rm{B}} \ga -19$ CADIS galaxies using FORS2 at the VLT and DOLORES at TNG we obtained oxygen abundances of 5  galaxies at $z \sim 0.4$ and 10  galaxies at $z \sim 0.64$.
Combining these measurements  with  published oxygen abundances of galaxies with $\rm{M}_{\rm{B}} \la -19$ 
we find evidence that a metallicity-luminosity relation   exists at medium redshift, but it is displaced to lower abundances and higher luminosities compared to the  metallicity-luminosity relation in the local universe.
Comparing the observed metallicities and luminosities of galaxies at  $z \la  3$ with P\'egase2 chemical evolution models we have found a favoured scenario in which 
the metallicity of  galaxies  increases  by a factor of $\sim 2$ between  $z \sim 0.7$ and today, and their luminosity decreases  by $\sim 0.5-0.9$\,mag.
 }

\titlerunning{Oxygen abundances of galaxies at medium redshift}
\authorrunning{Maier et al.}
\maketitle

%

\section{Introduction}

In the local universe, metallicity is well correlated with the
absolute luminosity (stellar mass) of galaxies
(Skillman et al. \cite{skilm}, Zaritsky et al. \cite{zarit}, Richer \& McCall \cite{richer},
Garnett et al. \cite{garnet}, Hunter \& Hoffman \cite{hunter}, Melbourne 
\& Salzer \cite{melb}, among others)
in the sense that more luminous
galaxies tend to be more metal rich.
This metallicity-luminosity relationship is often attributed to the
action of galactic superwinds: massive (more luminous) galaxies reach higher
metallicities because they have deeper gravitational potentials which
are better
able to retain their gas against the building thermal pressures from
supernovae, whereas low-mass systems eject  large amounts of metal-enriched gas  by supernovae-driven winds before high
metallicities are attained (e.g., MacLow \& Ferrara \cite{maclow}).
However,  a greater degree of gas consumption in luminous galaxies and/or suppressed infall of low-metallicity or pristine gas may also play a major role (Pagel \cite{pagel97}).

Optical emission lines from \hii\, regions have long been the primary
mean of gas-phase chemical diagnosis in galaxies (see, e.g., the
review by Shields  \cite{shields}).
The advent of large  telescopes and sensitive spectrographs has enabled direct
measurement of the chemical properties in the ionized gas of
cosmologically--distant galaxies with the same nebular analysis
techniques used in local \hii\ regions.
Distant galaxies subtend small angles on the sky, comparable to
typical slit widths in a spectrograph. 
A typical ground-based resolution element of 1\farcs0 corresponding
to a linear size of $\sim6$\,kpc at $z=0.5$  encompasses the entire
galaxy at medium redshift; so galaxy spectra tend to be integrated
spectra.
The use of {\it spatially-integrated} emission
line spectroscopy for studying the chemical properties of star-forming
galaxies at earlier epochs has been explored by Kobulnicky et al. (\cite{kob99}).
They concluded that spatially-integrated  emission line spectra
can reliably indicate the chemical properties of distant star-forming
galaxies. They also found  that, given spectra with sufficient signal-to-noise, the
oxygen abundance of these galaxies can be measured to within $\pm
0.2$\,dex using the \R23\ method, first suggested by Pagel et
al. (\cite{pagel}). The  \R23\ method allows to derive
metallicities
also for faint galaxies where only prominent emission lines are
observable (see more details in Sect. 4).

At medium redshift, all previous attempts to study the evolution of metallicity with cosmic time 
(Kobulnicky \& Zaritzky \cite{kobzar}, Hammer et al. \cite{hammer},  Carollo \&
Lilly \cite{carol}, Contini et al. \cite{contini}) have been based on 
continuum selected samples, the emission lines of which were identified using 
spectroscopy.
Owing to magnitude-limited selection effects in these surveys, the
samples of medium redshift galaxies   for which metallicities were
determined are biased towards luminous galaxies. 
 Thus, for instance, the CFRS sample used by Carollo \& Lilly 
(2001),  selected by  $I <22.2$, contains only bright
galaxies ($M_{B}<-20$) at 
redshift $0.6 \la z \la 1$, the continuum of which is dominated by an evolved stellar 
population.

Kobulnicky \& Koo (\cite{kobko}) found that star-forming Lyman-break galaxies at $z \sim
3$ are 2$-$4\,mag more luminous than local spiral galaxies of similar
metallicity, and thus are offset from the local luminosity-metallicity relation.
Also Pettini et al. (\cite{pettini})  found that Lyman break galaxies at $z\sim 3$
are significantly overluminous for their metallicities.

It seems therefore possible that the whole
metallicity-luminosity relation is displaced to
lower abundances at high or medium redshifts,
ilustrating how galaxies participate in the chemical evolution process.

The \textbf{C}alar \textbf{A}lto \textbf{D}eep \textbf{I}maging
 \textbf{S}urvey (CADIS, Meisenheimer et al. \cite{meise98},
 \cite{meise03}) allows the selection of galaxies   by their emission line fluxes detected via 
narrow-band (Fabry-Perot) imaging, regardless of their continuum
brightness. Thus, CADIS can reach galaxies with very low continuum brightness (and
absolute luminosity), i.e.  galaxies
which have  such faint continuum that they are only detectable by
their emission lines.
Combining metallicities for such faint CADIS galaxies in
 several narrow redshift bins with
 published abundances for more luminous galaxies at medium $z$  we aim
 to establish the metallicity-luminosity relation  for several look-back time bins in the range $0.3<z<1$.
The metallicity-luminosity relation at different reshifts is a valuable  probe of chemical galaxy evolution, and  a consistency check for chemical evolution models. 

This paper is structured  in the following way: In Sect. 2 we
describe how we select  galaxies at $0.39 \la z \la 0.65$ from the
CADIS emission line sample in order
to determine their metallicities by follow-up spectroscopy.
In Sect. 3 we present the spectroscopic follow-up of such emission
line galaxies. In Sect. 4 we compute oxygen abundances of the
spectroscopic sample.
Finally, in Sect. 5  we discuss the evolution of the
metallicity-luminosity relation with redshift.
The ``concordance'' cosmology with  $\rm{H}_{0}=70$ \kmsmpc,
$\Omega_{0}=0.3$, $\Omega_{\Lambda}=0.7$ is used throughout this
paper. For the solar oxygen abundance we use the value
 $12+\rm{log}(\rm{O/H})=8.87$ from Grevesse et al. (\cite{grevesse}).
Note that ``metallicity'' and ``abundance''  normally denotes ``oxygen
  abundance'' throughout this paper.


\section{Selection of emission line  galaxies  at medium redshift from CADIS}

CADIS
 combines  a moderately deep
multi-band survey (10 $\sigma$ limit, R$_{\rm{lim}} = 24$) with a deep
emission line survey employing an imaging
Fabry-Perot-Interferometer ($\rm{F}_{\rm{lim}} = 3 \cdot 10 ^ {-20}
 \rm{W}  \rm{m}^{-2}$), which covers
three waveband windows (henceforth called FP windows)  essentially free of  atmospheric OH emission lines:  A\,($\lambda = 702 \pm 6$\,nm),
B\,($\lambda = 819 \pm 5$\,nm), and C\,($\lambda = 920 \pm 8$\,nm).
An emission line detected in the Fabry-Perot scan can be any of the prominent
nebular lines, e.g., \lya,  \OII,  \Hb, \OIIIa, or \Ha.
The line identification is done by means of the line profile, by fitting
template galaxy spectra to the observed multi-filter SED, and  
by considering the flux in
``veto-filters'' at the expected wavelengths of other lines (Hippelein
et al. \cite{hippe}, Meisenheimer et al \cite{meise03}).

\begin{figure*}[ht!]
\centerline{\psfig{figure=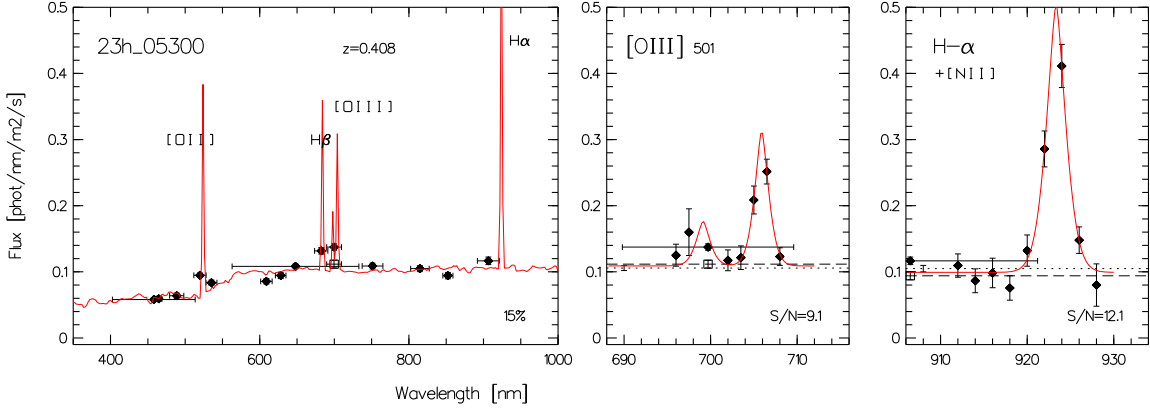,width=16cm,angle=270,clip=true}}
\caption
[The galaxy 23h-671455 at $z\approx 0.41$]
{\label{spcad5300} \footnotesize The galaxy 23h-671455 at $z \approx 0.41$: Photometry
in all 14 optical
CADIS filters fitted by a
  continuum-model with overlayed prominent emission lines (left panel), the Fabry-Perot measurements in window A with a
  [OIII] doublet profile fitted to the observed flux data (center), and
  the Fabry-Perot measurements in window C with a \Ha\,  $+$ 
  \NIIdob\, three-line profile fitted
  to the observed flux data (right panel).
  Note that the \OII\,
    line is seen in the veto-filter at 522\,nm.
 The numbers in the lower right edges of the right panels specify the 
signal-to-noise ratios, $F_{line}/\sigma_{line}$; 
 $F_{line}$  and $\sigma_{line}$ are derived by fitting
the Fabry-Perot instrument profile to the observed flux data points;
see also Hippelein et al. (\cite{hippe}).
}
\end{figure*}

\begin{figure*}[ht!]
\centerline{\psfig{figure=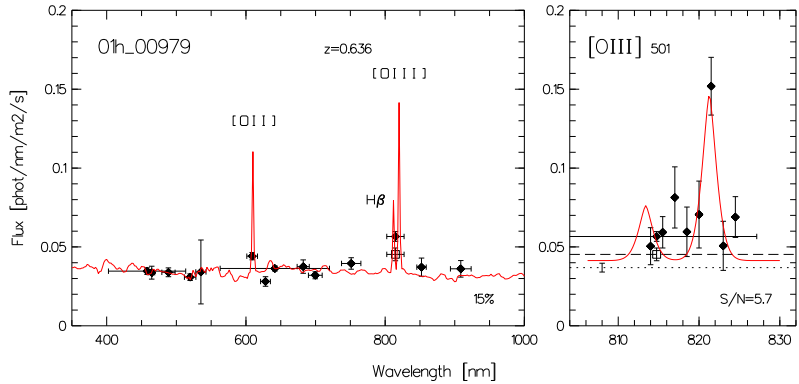,width=11cm,angle=270,clip=true}}
\caption
[The galaxy 01h-584247 at $z \approx 0.64$]
{\label{spcad979} \footnotesize The galaxy 01h-584247 at $z \approx 0.64$: Photometry
in all 14 optical
CADIS filters fitted by a
  continuum-model with overlayed prominent emission lines (left panel), and the Fabry-Perot measurements in window B with a
  [OIII] doublet profile fitted to the observed flux data (right
  panel).
 Note that the \OII\,
    line is seen in the veto-filter at 611\,nm.}
\end{figure*}

The analysis of FP window A and B is nearly completed in four of the 6
CADIS fields with FP observations (Table 1 in Maier et al. \cite{maier03}). In these four CADIS fields we have found 128  galaxies   by  their \OIIIa\, emission  detected in the FP windows A
or B; they have redshifts  in the redshift bin $0.392<z<0.415$, and $0.625<z<0.648$,  respectively.
Figures \ref{spcad5300} and \ref{spcad979} show two examples of
    galaxies detected by their \OIIIa\, emission line in the
    FP scan.
The \OIIIa\, line of the galaxy 23h-671455 (Fig.\,\ref{spcad5300}) 
 is detected in the FP  window A, and we see the \OII\, emission line in the 
veto-filter at 522\,nm. Moreover, \Ha\, and the \NIIdob\, doublet
are seen in window C.
For the galaxy 01h-584247 (Fig.\,\ref{spcad979}), the \OIIIa\,
line is seen by the Fabry-Perot in window B, and we detect the \OII\,
    line in the veto-filter at 611\,nm.

 From the 
sample of 128 objects detected by their \OIIIa\,   line 
in FP windows A or B we selected  emission line galaxies for
spectroscopic-follow up in  order to determine their metal abundances.
Oxygen lines are the most
useful for measuring the abundance, since all important ionization
stages can be observed.  The temperature-sensitive  \OIIItemp\ line
is often too weak to be measured even in the local universe
(\OIIItemp/\OIIIa\, $ \la  1/50 $).
To overcome this problem,  Pagel et al. (\cite{pagel}) suggested the empirical abundance indicator

\begin{equation}
  \label{r23rel}
\R23 = \frac{\rm{I}(\OII) + \rm{I}(\OIII)}{\rm{I}(\Hb)}
\end{equation}

for faint objects, such as galaxies at high redshift. The method has
been later refined by McGaugh
(\cite{mcgaugh}) using a set of photoionisation models, and
observationally confirmed to be reliable by
Kobulnicky et al. (\cite{kob99}).
For the 128 selected objects the \OIIIa\, line is measured in the respective Fabry-Perot window,
while \Hb\, must be measured by spectroscopic follow-up,
and \OII\, is detected by a veto filter at 522 or 611\,nm.
 However, since the determination of the \OII\,
flux  relies only on a medium band filter, this flux is often not accurate
enough  to determine oxygen abundances with the \R23\, relation.
Therefore, spectroscopic follow-up  is
necessary  to measure not only the \Hb\, line, but also \OII.
Since the \OIII\, and  \Hb\, lines lie close in wavelength to each other,
the accurate fluxes of the \OIII\, lines are  provided for free by the spectroscopic follow-up.

\begin{figure}[ht!]
\centerline{\psfig{figure=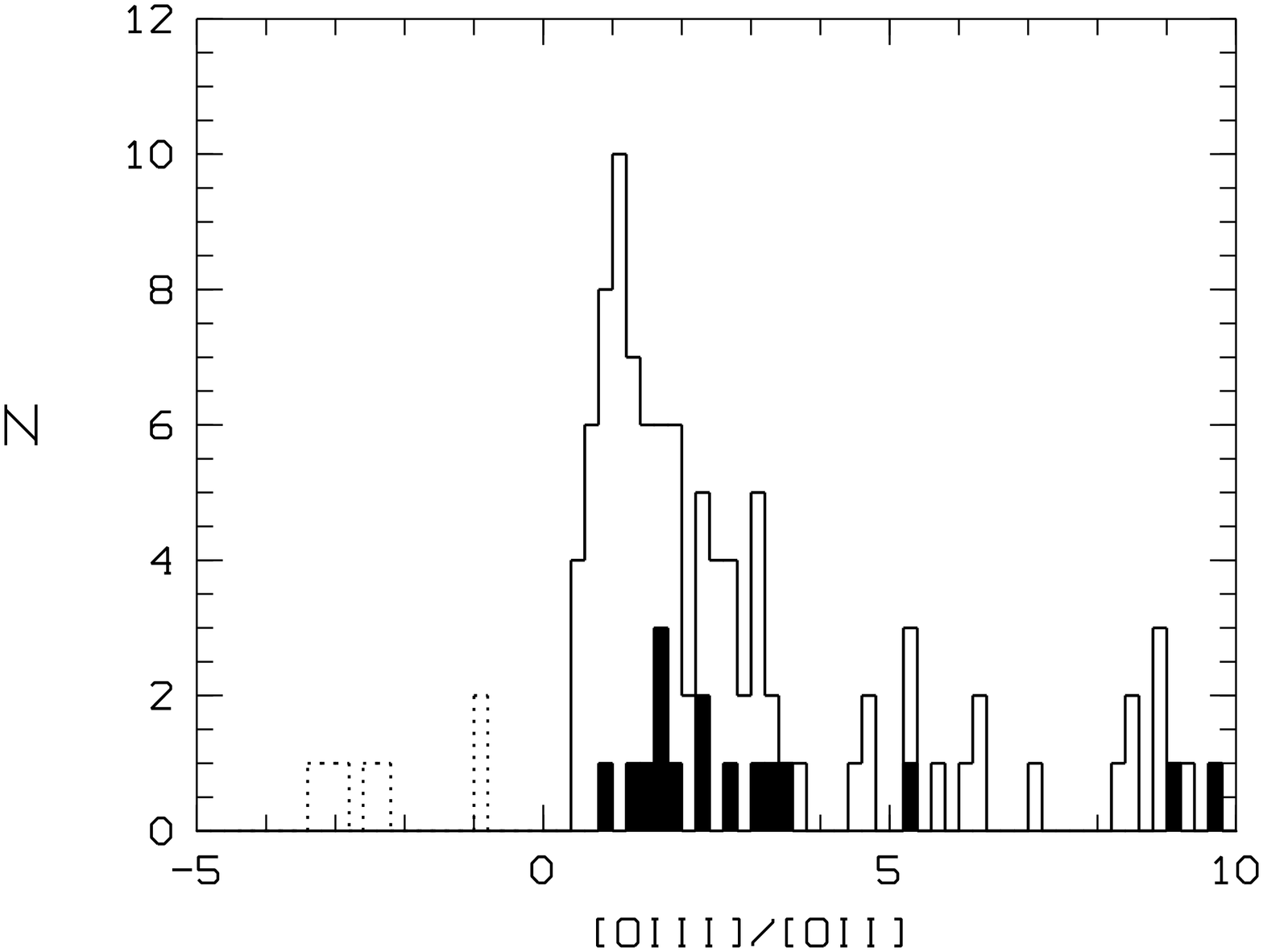,width=10cm,angle=270,clip=true}}
\caption
{\label{auswahloiii} \footnotesize The histogram of  F(\OIIIa)/F\OII)
 flux ratios (as  determined by the CADIS
filter and Fabry-Perot measurements) of  128 emission line galaxies selected by their
  \OIIIa\, line seen in the Fabry-Perot; the filled
boxes show the CADIS flux ratios for the galaxies with follow-up
 spectroscopy listed in Table \ref{elgals}. Note that the negative
 values (dotted lines)
 come from the fact that, for galaxies
 with undetectable \OII\, flux, the uncertainties in the
 interpolated continuum and the background noise place the formally
 measured F(\OII) flux symmetrically around F(\OII)$=0$. 
}
\end{figure}

From our 128 objects detected by their \OIIIa\, emission line seen in the
Fabry-Perot we obtained spectra of 21 objects with absolute magnitudes  fainter than $\rm{M}_{\rm{B}} \approx -19$,
    which were able to be placed on the slitmasks together with  Ly-$\alpha$
candidates (Maier et al. \cite{maier03}) in order to maximize the
    number of galaxies observed; see  Section 3 for the details of the observations.
The large wavelength coverage of the CADIS filters allows  to derive
precise absolute magnitudes without using any K-correction factor, since
the flux value at the restframe wavelength is calculated by linear
interpolation between the measurements of two adjacent filters.
In order to extend the sample of already  published
    metallicities of galaxies at medium redshift to fainter absolute magnitudes, we selected for the
    spectroscopic follow-up objects 
with $\rm{M}_{\rm{B}} \ga -19$.
We removed five
objects from this sample because the \Hb\, line had too low
signal-to-noise ratio for reliable oxygen abundance determinations
(see Kobulnicky et al. \cite{kob99} for the discussion of the
signal-to-noise required in order to get reliable metallicity measurements with
the \R23\, method).
The remaining 16 galaxies are listed in  Table \ref{elgals}.
As shown in Fig.\,\ref{auswahloiii}, this final sample
is only slightly biased
towards special  F(\OIIIa)/F(\OII) ratios --  a
ratio which  decreases with   metallicity (Stasinska \& Leitherer
\cite{stasinsk}) -- in the sense that objects with
F(\OIIIa)/F(\OII)\,$<1$ are underrepresented (28\% in the full
sample, 12\% in the spectroscopic sample).

\section{Spectroscopic observations}

The spectroscopic follow-up observations of  emission line galaxies in two redshift bins, at $0.392 \la z \la 0.415$, and at $0.625 \la z
\la 0.648$, in the CADIS 01h-
 and
23h-fields
were obtained in the summer and autumn of 2001, and in the summer of 2002,  using  FORS\,2 at the
 VLT.
 Galaxies
in 09h-field, and one galaxy in 01h-field  at $z\approx0.64$  (01h-5085)
 were observed
with the Low Resolution Spectrograph DOLORES at Telescopio Nazionale
Galileo (TNG).
 The slitmasks contained 1\farcs0 up to 1\farcs6   wide slits; the
length of the slits varied between 10\farcs0, and 20\farcs0. 
Since we found from our first observations that the 1\farcs0 slit may miss (part of)
the line emitting region (see discussion in  Maier et
al. \cite{maier03}), we used wider slits (up to 1\farcs6) for the
observations with TNG (masks \emph{e}, \emph{f}), and  for the VLT
observations in summer 2002 (masks \emph{d}, \emph{i}).

\begin{table*}[ht!]
\caption
{\label{observ} \footnotesize Spectroscopic follow-up observations.} 
\begin{center}

\begin{tabular}{c c c c c c c}
\hline
\hline
Field &  Mask       & t(sec) & Grism &Res.(nm/slit)&Slitwidth& Inst./Telescope   \\
\hline
01h   &   a  & 12860    &  300I  &1.2&1\arcsec.0&  FORS2/VLT\\
01h   &   b  & 12500    &  600RI &0.8&1\arcsec.0&  FORS2/VLT\\
01h   &   b  & 7500     &  600R  &0.5&1\arcsec.0&  FORS2/VLT\\
01h   &   c  & 6085     &  600RI &0.8&1\arcsec.0&  FORS2/VLT\\
01h   &   c  & 1700     &  600R  &0.5&1\arcsec.0&  FORS2/VLT\\
01h   &   d  & 15750    &  300I  &1.7&1\arcsec.4&  FORS2/VLT\\
01h   &   e  & 6000     &   LRR  &1.2&1\arcsec.1&  DOLORES/TNG\\
\hline
09h   &   f  & 15600    &   HRR  &0.5&1\arcsec.6&  DOLORES/TNG\\
09h   &   f  & 3000     &   MRB  &1.0&1\arcsec.6&  DOLORES/TNG\\
\hline
23h   &   g  & 16080    & 300I   &1.2&1\arcsec.0&  FORS2/VLT\\
23h   &   h  & 9000     & 600RI  &1.8&1\arcsec.0&  FORS2/VLT\\
23h   &   i  & 15750    & 300I   &1.7&1\arcsec.4&  FORS2/VLT\\
\hline  
\end{tabular}
\end{center}
\end{table*}

For the VLT observations  we used three different grisms,
the
lower resolution
300\,I grism, which gives a spectral resolution of about  1.2\,nm
at 800\,nm for 1\farcs0\, wide slitlets;
the 600\,RI grism, which gives a higher spectral resolution of about
0.8\,nm  at 800\,nm
 for 1\farcs0\, wide slitlets, and the 600R grism, which  gives a spectral resolution of about
 0.5\,nm  FWHM at 600\,nm.
 However, since the objects have to be put on the masks according
 to the wavelength range to be observed, the 600RI grism has a
 smaller  spectroscopically field-of-view than 
 the 300\,I grism, and thus allows less
 freedom in target positions. 
 Moreover,  one can measure
 only  the \OIII\, and \Hb\, lines of
a galaxy at $z \approx 0.64$, and,
in order to get \OII,   the 600R grism  has to be used additionally. 

Spectra were reduced using the package LONG provided by MIDAS.
Images were bias-subtracted first.
In order to
remove the low spatial frequencies along the dispersion axis from the
flatfield, we averaged the original image along the slit, fitted
the resulting one-dimensional image by a polynomial of second order,
and divided the original flatfield by this image to obtain a
normalized flatfield.
Every mask contained about 20 single spectra. 
Every single spectra was extracted,
treated as a long slit spectra, and divided through the respective normalized
flatfield.
For the wavelength calibration night sky lines were used, which
have the advantage to be taken with the telescope and spectrograph in
the same orientation as the science spectra.
The one-dimensional spectra of each object were extracted with an
 aperture of about 10 pixels, i.e., about 2\arcsec, using the algorithm by Horne (\cite{horne}).
Fluxes were calibrated from digital units (ADUs) to physical flux
units (W m$^{-2}$ s$^{-1}$ \am$^{-1}$) using 
multiple observations of the spectrophotometric standard stars LTT
7379, EG 274, and LTT 7987 (Hamuy et al. \cite{hamuy92}, \cite{hamuy94}).

For the  observations with TNG we used the HRR and MRB grisms for 
galaxies at $z\approx0.4$, and the LRR grism for the galaxy 01h-5085
at $z\sim 0.64$.
Data reduction of the spectra was performed similar to the VLT data reduction.
For the flux calibration the spectrophotometric standard star HD93521
(Oke \cite{oke}) was used.
Table \ref{observ} lists the details of the observations with the VLT and TNG.

\section{Flux ratios and oxygen abundances}

\subsection{Emission line measurements}

Emission line fluxes  were measured interactively using the integrate/line
routine in MIDAS.
The flux error is calculated as $\sqrt{N} \times \sigma$, where $N$ is
the number of pixels summed in a given emission line, and $\sigma$ is
the root-mean-squared variation in a region of the spectrum next to
the emission line.
This yields typical errors of 5\% for \OIIIa, \OIIIb, and \OII, and
5\% to 12 \%  for \Hb, depending on the
position of \Hb\, between the night sky lines..
The error of \R23\, is thus dominated by \Hb. 
Flux ratios were corrected for underlying Balmer
absorption
with a general \Hb\, absorption equivalent width of 2\am, representative of local irregular,
HII, and spiral galaxies (see e.g., McCall et al. \cite{mccall},
Skillman  \& Kennicutt \cite{skillm93}, Izotov et al. \cite{izo94}),
and in agreement with the newest results for \Hb\, absorption
  equivalent widths   from  the study of about 600 SDSS strong emisison
  line galaxies (Kniazev et al. \cite{kniazev}).
 
\subsection{Extinction}

The dereddened  value for a flux line ratio,
$\rm{I}(\lambda 1)/\rm{I}(\lambda 2)$, is given by:

\begin{equation}
\frac{\rm{I}(\lambda 1)}{\rm{I}(\lambda 2)} = \frac{\rm{F}(\lambda
    1)}{\rm{F}(\lambda 2)} \times 10^{c(f(\lambda 1)- f(\lambda 2))},
\end{equation}  

where $\rm{F}(\lambda)$ is the observed flux at a given wavelength, c is the logarithmic
reddening parameter that describes the amount of reddening relative to
\Hb, and $f(\lambda)$ is the wavelength-dependent reddening function (Whitford   \cite{whitford}).
The value of c can be extimated  from the relation $c = 1.47 \times \rm{E}_{\rm{B-V}}$ from Seaton (\cite{seaton}).

The ratio $\rm{I}(\OIIIa)/\rm{I}(\Hb)$ which goes into the \R23\,
relation is:

\begin{equation}
\frac{\rm{I}(\OIIIa)}{\rm{I}(\Hb)} =
\frac{\rm{F}(\OIIIa)}{\rm{F}(\Hb)} \times 10^{-0.034\,c}.
\end{equation}  

Thus, since the
\OIII\, and \Hb\, lines lie close in wavelength to each
other, the extinction is small and can be neglected for the
$\rm{I}(\OIIIa)/\rm{I}(\Hb)$ ratio (the same applies also to the
$\rm{I}(\OIIIb)/\rm{I}(\Hb)$ ratio).

The other ratio, $\rm{I}(\OII)/\rm{I}(\Hb)$, which goes into the \R23\,
relation is

\begin{eqnarray}
\label{aextink}
  \frac{\rm{I}(\OII)}{\rm{I}(\Hb)} & = &
\frac{\rm{F}(\OII)}{\rm{F}(\Hb)} \times 10^{0.308\,c} \nonumber\\
& & =\frac{\rm{F}(\OII)}{\rm{F}(\Hb)} \times  \alpha,
\end{eqnarray}

with $\alpha =  10^{0.308\,c}$.
Thus,  the effect of reddening in the \R23\, expression can be written as:

\begin{equation}
 \label{r23extink} 
\R23\,  = 
 \frac{\alpha \times \rm{F}(\OII) + \rm{F}(\OIII)}{\rm{F}(\Hb)}.
\end{equation}

 Assuming  case B Balmer recombination, 
with a temperature of 10\,000\,K, and  a density of 100\,${\rm{cm}}^{-3}$
(Brocklehust \cite{brockle}), the predicted  dereddened intensity ratio of \Ha\, to \Hb\,
is 2.86 (Osterbrock \cite{osterb}).
The effect of reddening on the ratio \Ha/\Hb\, can be written as
$\rm{I}(\Ha)/\rm{I}(\Hb) =  \rm{F}(\Ha)/\rm{F}(\Hb)\times
10^{(-0.332)c}$.
Using the spectroscopic follow-up measurements of the \Ha\,
line for three galaxies at $z \approx 0.4$  (see, e.g.,
Fig. \ref{hanii}), 
and the CADIS Fabry-Perot measurements of the
\Ha\, line available in window C for   galaxies at $z
\approx 0.4$   (see  Fig. \ref{spcad5300}),
the values of $c$ for the 5 galaxies at  $z \approx 0.4$   are of
order of $0 - 0.1$, yielding a value of $\alpha$ in the range 1 to 1.07.
This means that 
the extinction  (see equation \ref{r23extink}) can  be neglected  with
respect to the errors for the faint emission line
galaxies  ($\rm{M}_{\rm{B}} \ga -19$) at $z \approx 0.4$, when
calculating \R23.
We assume that the same is true 
for galaxies with  $\rm{M}_{\rm{B}} \ga -19$ in the redshift bin  $0.625<z<0.645$. 

This assumption is supported by the fact that the \OIIIa\, line flux is larger than  the \OII\, line flux for
the 11  galaxies at $z\approx0.64$  of our sample (see Table
\ref{elgals}). Therefore, even allowing for a wider range
$1<\alpha<2$ and typical large F(\OIIIa)/F(\OII) ratios,  
equation \ref{r23extink} shows
that  the  \OIII\, and \Hb\, measurement errors, and not  the unknown
value of $\alpha$, dominate the error of the \R23\, ratio. Therefore, we do not
correct the line fluxes for
extinction for the galaxies in our sample.

\subsection{The \R23 relation}
\label{r23rel}

Because the temperature-sensitive line \OIIItemp\, is not detectable in the
spectra of galaxies at $z\ga0.4$,
the oxygen abundance has to be determined using the empirical
calibration between oxygen abundance and \R23.  Pagel et al. (\cite{pagel})
noted, based on a sample of extragalactic H\,II regions, that the
measured T$_{e}$, O/H and \R23 were all correlated.
This works because of the relationship between O/H and nebular cooling:
For metal-rich regions, the cooling  in the ionized gas is dominated by emission in IR
fine-structure lines (primarily the [O\,III] 52 $\mu$m and 88
$\mu$m lines), so as O/H increases, the nebula becomes cooler.
In response to that, the highly excited optical forbidden lines,  the 
[O\,III] lines, become weaker as O/H increases (excitation goes
down as T decreases). At very high abundances,
$12+\rm{log}(\rm{O/H}) \ga 
8.7$, \hii\, regions are very cold ($T_{e} < 6000$\,K) and
values of \R23\, are small.
For $12+$log (O/H) $\la8.2$, the relation between \R23 and O/H reverses, such
that \R23 decreases with decreasing abundance.
This occurs because at very low metallicities the IR fine-structure
lines no longer dominate the cooling because of the lack of  heavy
elements. As a result, the forbidden lines  reflect the
abundance in the gas almost in a proportional way.

\begin{figure}[h]
\centerline{\psfig{figure=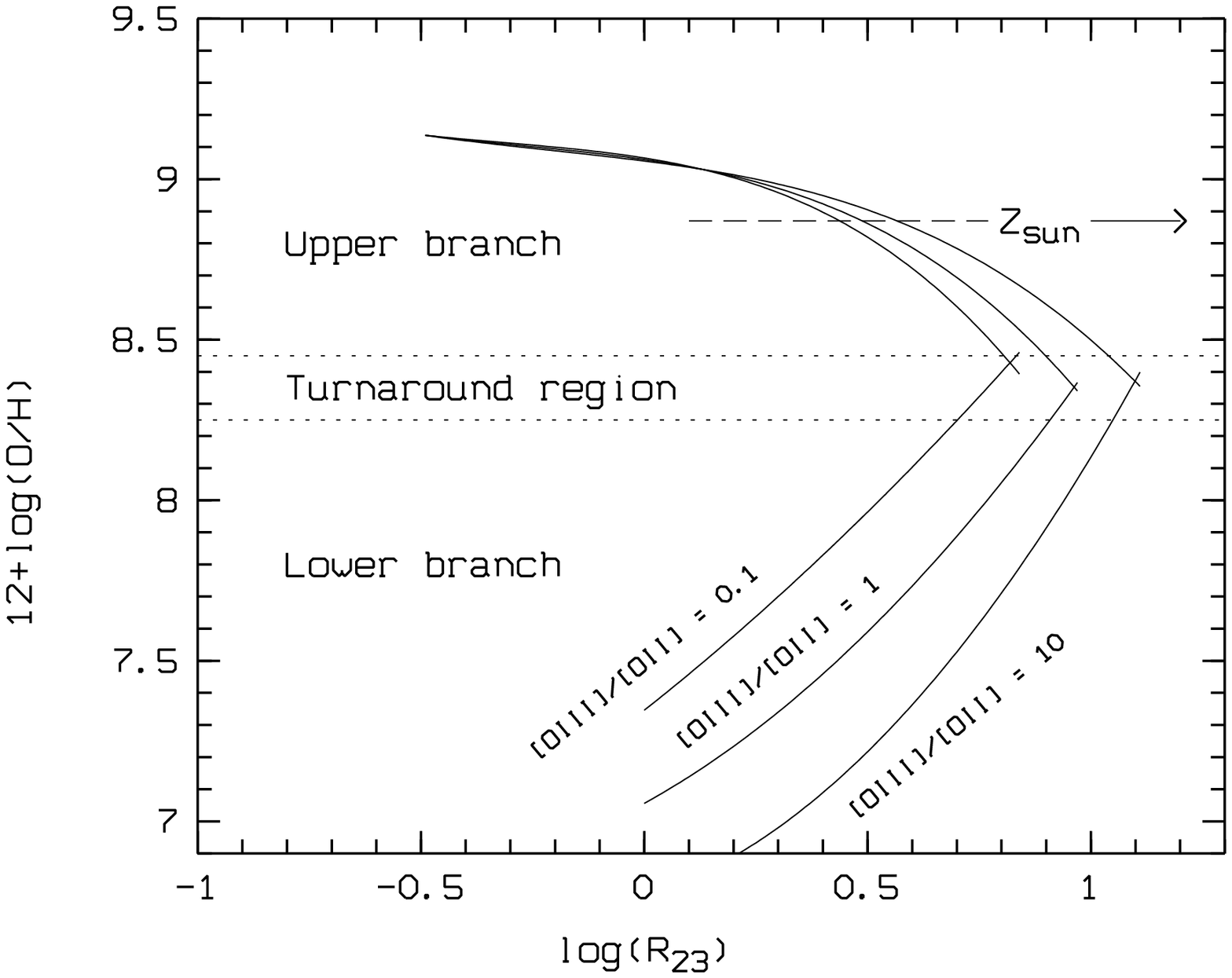,width=9cm,angle=270}}
\caption
[The oxygen abundance 12+log(O/H) as a function of the line ratio
$\rm{log}\,\rm{R}_{23}$]
{\label{gaugh} \footnotesize 
The oxygen abundance 12+log(O/H) as a function of the line ratio
$\rm{log}\,\rm{R}_{23}$.
The calibration between $\rm{R}_{23}$ and O/H (solid lines)
 using the models from McGaugh (\cite{mcgaugh}) shows the effect of varying the
 ionization parameter in terms of the observable line ratio
 \OIII/\OII\,.
 On the lower, metal-poor branch, the ionization parameter becomes
important.
The ratio [O\,III]/[O\,II] of 10, 1.0, and 0.1 correspond (very roughly) to
ionization parameters U of $10^{-1}, 10^{-2}$, and $10^{-4}$.
 The two  horizontal dotted lines separate the different regions
 of oxygen abundances.
 The  arrow and the  horizontal dashed line indicate the solar oxygen abundance of
 $12+\rm{log}(\rm{O/H})=8.87$ (Grevesse et al. \cite{grevesse}).
 }
\end{figure}

As a consequence, the \R23\, indicator is  not a monotonic function of oxygen
abundance (see Figure \ref{gaugh}). At a given value of \R23, there are two possible choices of
the oxygen abundance. Roughly, the low abundance or ``lower branch'' is
defined by  $12+\rm{log}(\rm{O/H}) \la 8.2$, whereas the high abundance or
``upper branch'' is defined by $12+\rm{log}(\rm{O/H}) \ga  8.5$.
Therefore, an
additional indicator (e.g., the \NII\, line) is  needed in order
to resolve the degeneracy in \R23.
About 50\% of the emission line galaxies in our sample have log\R23 in the range $0.8-1$ placing these galaxies in the ``turnaround'' region $8.2 <  12+\rm{log}(\rm{O/H})  <  8.5$ of the oxygen abundance versus \R23\, diagram (see Fig.\,\ref{gaugh}).
For the remaining galaxies
we describe in the following how  the decision, whether a galaxy lies on
the upper or on the lower branch,  has been taken.

\subsubsection{Breaking the \R23\, degeneracy by the \NII/\Ha\ ratio}

The \NII/\Ha\, line ratio can be used  to break the degeneracy
of the \R23\, relation, if one is able to  measure and  separate
\NII\, from \Ha.
Denicolo et al. (\cite{denicolo})  calibrated the
N2 estimator, defined as N2\,$=\rm{log}(I(\NII)/I(\Ha))$, vs. the oxygen
abundance,
 using a sample of \hii\, galaxies
having
accurate oxygen abundances, plus photoionization models
covering a wide range of abundances.

\begin{figure*}[ht!]
\centerline{\psfig{figure=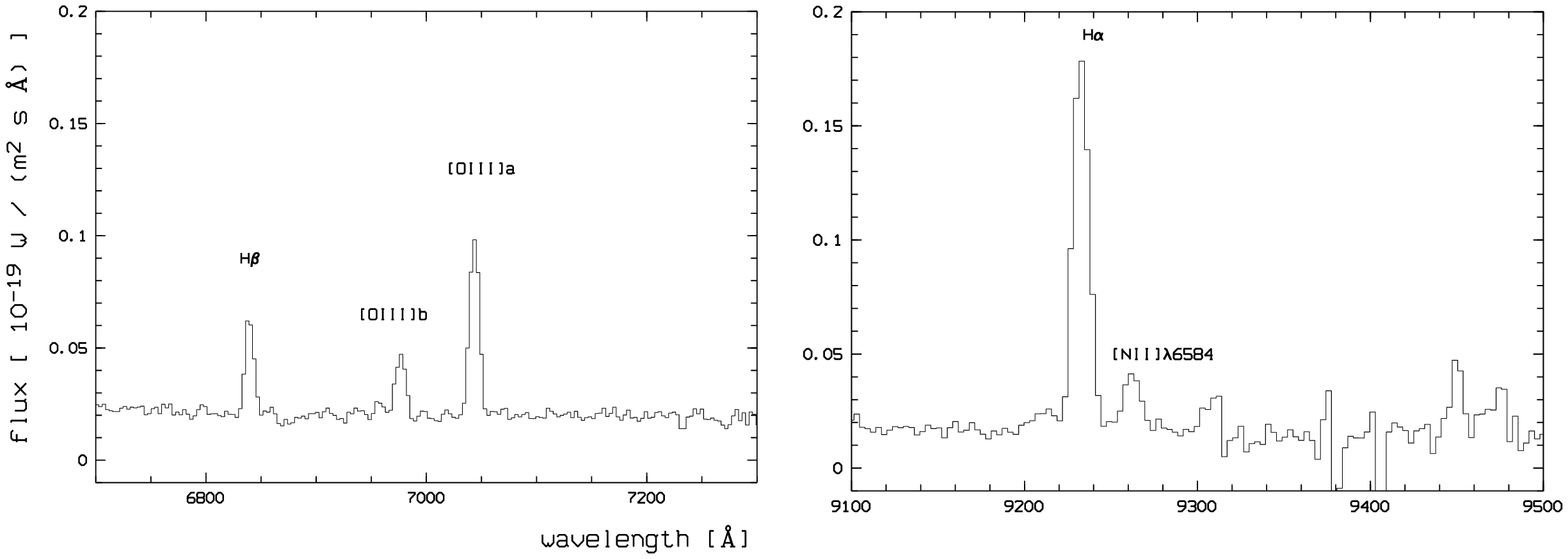,clip=t,width=12cm,angle=270}}\vspace*{1mm}
\centerline{\psfig{figure=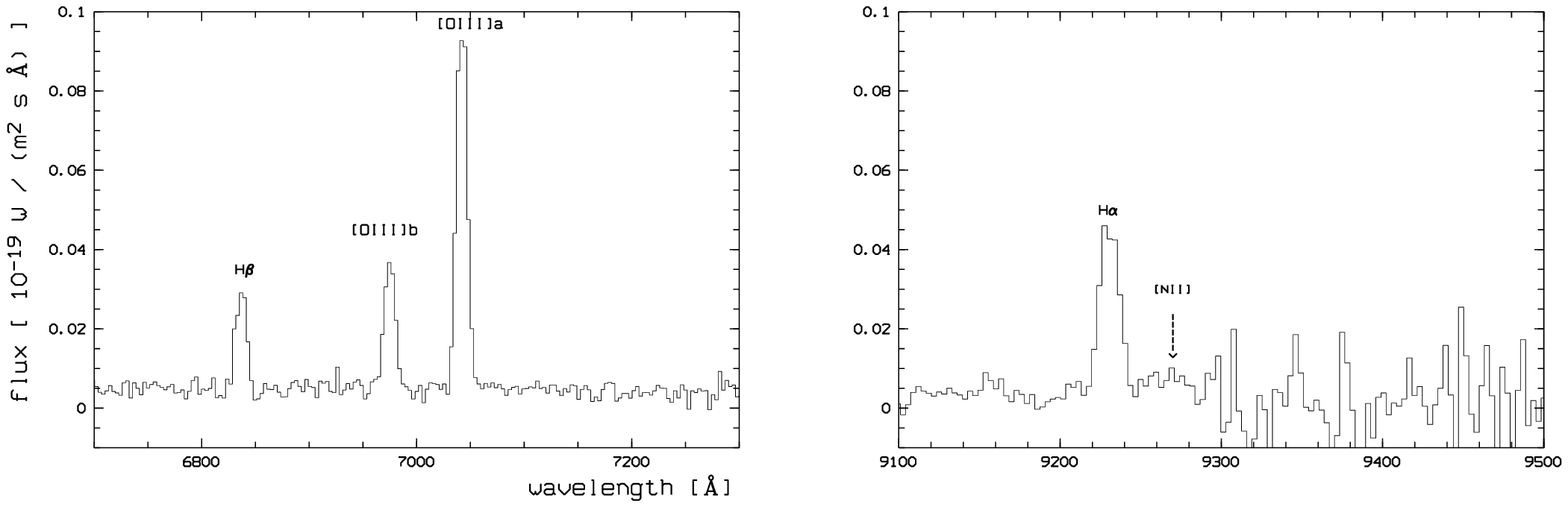,clip=t,width=12cm,angle=270}}
\caption[VLT Spectra using the 300I grism with
  FORS2.]
{\label{hanii} \footnotesize VLT Spectra (mask \emph{i}) using the 300I grism with
  FORS2. The ``features'' observed in the continuum are largely due to
  residuals from the subtraction of strong OH emission lines.
  
\textbf{Upper panel:} The galaxy 23h-671455
  (M$_{\rm{B}}=-19.0$) at $z \approx 0.4$
  shows a \NII\,/\Ha\, line ratio of  $\sim0.1$, and lies in the
  turnaround region of the \R23\, relation.

\textbf{Lower panel:}
The galaxy 23h-683506 (M$_{\rm{B}}=-17.3$) at $z \approx 0.4$
  shows a \NII\,/\Ha\, line ratio $<0.1$, placing this
galaxy on the lower branch of the \R23\, relation.

}
\end{figure*}

When the secondary production of
nitrogen dominates, at somewhat higher metallicity, the  line ratio \NII/\Ha\,
increases with oxygen abundance.
At very low metallicity, N2 scales simply as the nitrogen
abundance to first order. However,  in this
metallicity regime, the nitrogen abundance shows a large  scatter
relative to the oxygen abundance, since the nitrogen abundance is much
more sensitive to the history of star formation in the galaxy considered.
As a result, N2 is probably not very useful to estimate oxygen
abundance except as a means of determining the  branch for the 
application of the \R23\, method. The division between the upper and
the lower branch of the \R23\, relation occurs around
\NII\,/\Ha\, $\sim0.1$.
\NII\,/\Ha\, line ratios were  measured by follow-up spectroscopy for  three
CADIS galaxies at $z\approx0.4$. 09h-448495 has an \NII\,/\Ha\, line ratio $>0.1$, placing this
galaxy on the upper branch of the \R23\, relation; the galaxy
23h-671455 has a \NII\,/\Ha\, line ratio of $\sim0.1$ and lies in the
turnaround region;
and 23h-683506  has a \NII\,/\Ha\, line ratio $<0.1$, placing this
galaxy on the lower branch of the \R23\, relation (see spectrum in  Figure
\ref{hanii}, lower panel).

\begin{figure}[ht!]
\centerline{\psfig{figure=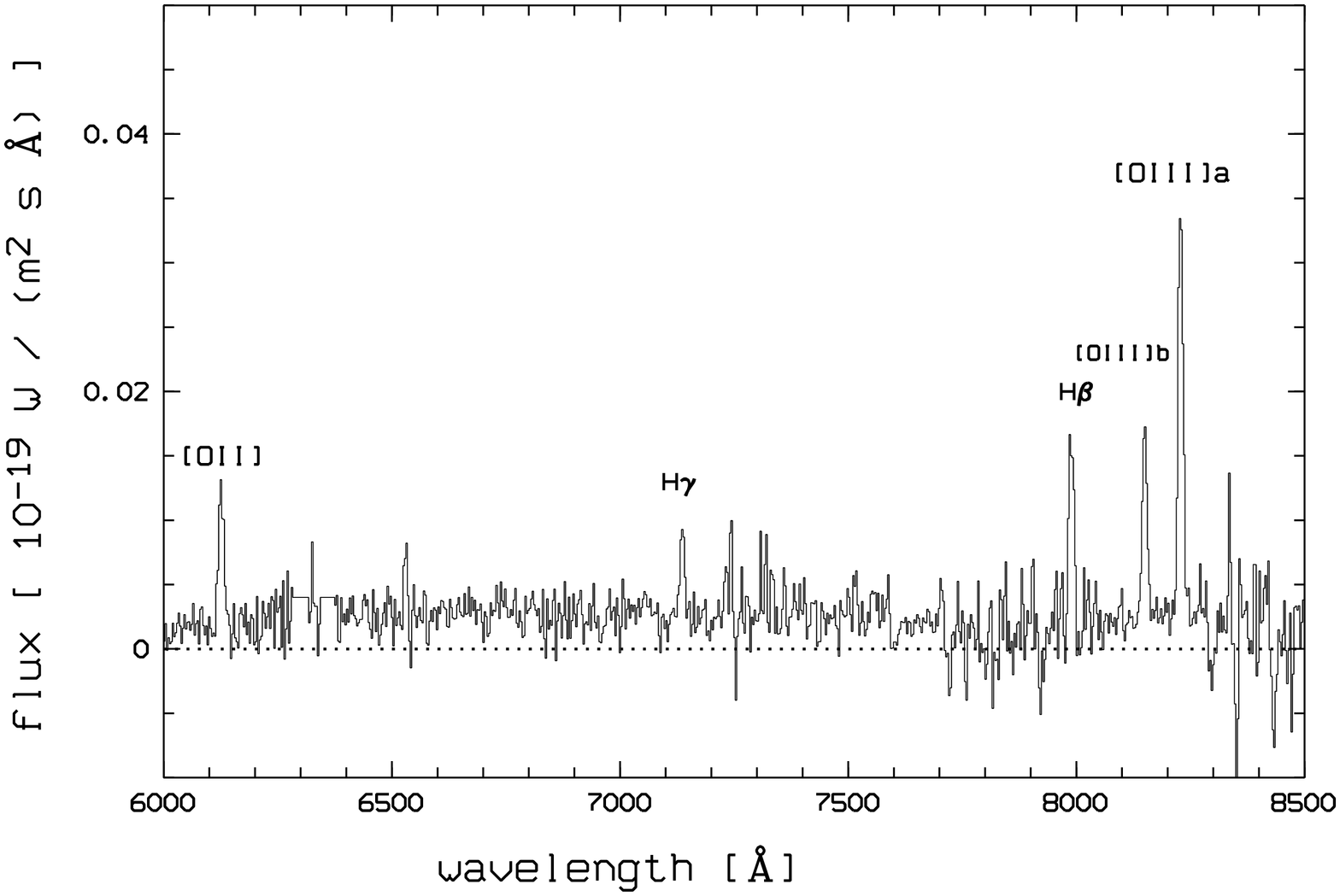,clip=t,width=10cm,angle=270}}\vspace*{1mm}
\centerline{\psfig{figure=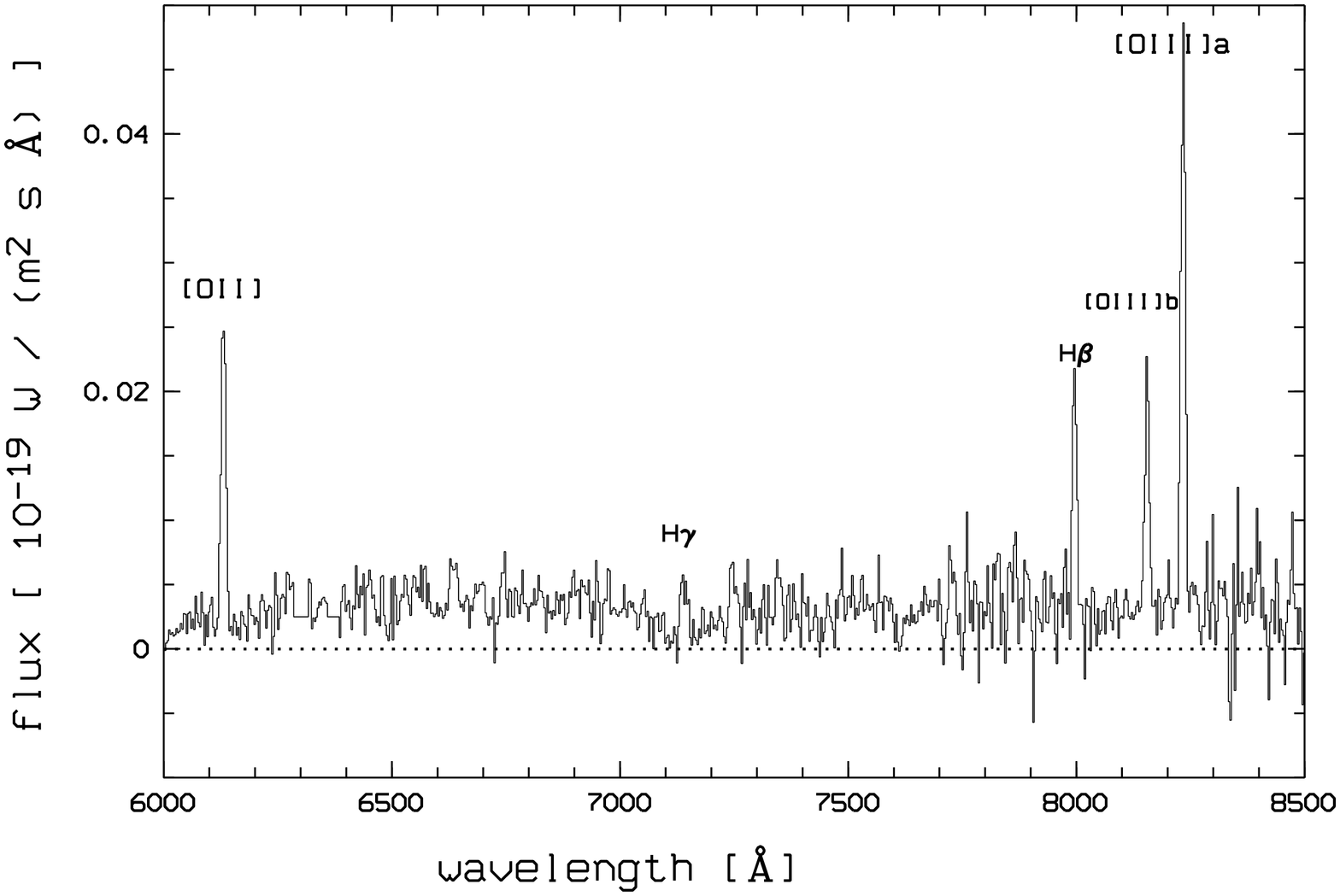,clip=t,width=10cm,angle=270}}\caption[VLT Spectra using the 300I grism with
  FORS2.]
{\label{obj5159} \footnotesize VLT Spectra using the 300I grism with
  FORS2 of
two galaxies at $z \approx 0.64$: 23h-5159 with M$_{\rm{B}}=-17.5$
(upper panel), and 23h-6512 with M$_{\rm{B}}=-18.3$ (lower panel).
The ``features'' observed in the continuum are largely due to
  residuals from the subtraction of strong OH emission lines.
}
\end{figure}

\subsubsection{Breaking the \R23\, degeneracy by high \OIIIa/\OII\, flux ratios}

From the remaining galaxies without measured \Ha\, and \NII\, (which
would require near-infrared spectroscopy for $z\sim 0.64$)
three galaxies, 09h-542442, 01h-246610, and 23h-408246,  show a high \OIIIa\,
to \OII\, flux ratio (greater than five).
If we put these galaxies on the upper branch, we would get
metallicities of $0.5\,Z_{\sun}<Z<1Z_{\sun}$.
There are no galaxies in the local universe in this high metallicity range
which show a such high  \OIIIa/\OII\, ratio. Assuming that the physical
properties of the interstellar medium are the same in the local
universe and at medium redshift, these galaxies cannot have a such high
metallicity.
Therefore, we have to put them on the lower branch. 
We used this criterion also for the galaxies 23h-441445 and 23h-552534
(spectra shown in  Figure
\ref{obj5159}).

 We exclude  the galaxy 01h-628521  oxygen
 abundance (indicated in Table \ref{elgals} as (?)) from the
following discussion of oxygen abundances, since we cannot be sure on which
 branch of the \R23\, relation to place it (\OIIIa/\OII\,$\sim1$). Near infrared spectroscopy will be required in order to measure the
\Ha\, and \NII\, lines of this galaxy, and to determine on which
branch of the \R23\, relation it has to be placed.

\subsection{Oxygen abundances}

We use the most recent \R23\, formulation by Kobulnicky et al. (\cite{kob99})
of the analytical expressions by McGaugh (\cite{mcgaugh})  to
determine the oxygen abundances of the emission line galaxies in our sample. These
formulae express the oxygen abundance, $12+$log\,(O/H), in terms of
\R23\, and the ionization index \OIII/\OII. The branch of the \R23\,
relation was selected as described in section \ref{r23rel}.
The computed oxygen abundances are shown in Table \ref{elgals}.
Additional to the measurement uncertainties
of the emission line ratios,  an uncertainty of $\sim 0.12$\,dex in the
oxygen abundance has been assumed, which comes from uncertainties in
the photoionization models and ionization parameter corrections for
the \R23\, method. 

\begin{table*}[ht!]
\caption{\label{elgals} Emission line galaxies with unreddened line
 strengths and computed \R23\, ratios.
 $^{a}$ to $^{i}$ indicates the mask with which the object was
 observed (see Table \ref{observ}). Column \emph{Branch} indicates if the object
 lies on the lower (L) or on the upper (U) branch of the O/H versus
 \R23\, relationship,  falls in the turnaround region (T), or if no
 decision between  the upper (U) and lower branch can be made (?).
 The table is ordered in bins of increasing 
 redshift, at $0.392 < z < 0.415$, at $ 0.625 < z < 0.648$, and $z>0.68$.
}
\begin{center}

\begin{tabular}{c c c c c c c c c c}
\hline
\hline
\#   & $z_{spec}$ & $M_{B}$ & [OIII]/[OII] & [OIII]/\Hb & log\R23 &
    Branch & O/H$_{L}$  &
O/$H_{U}$ &  $12+$log(O/H)\\
\hline
09h-542442$^{f}$ & 0.3985 & $-$18.1 & $11.11\pm 0.79$ & $3.03\pm 0.34$ &
$0.77\pm 0.10$  & L &7.64 & 8.74  &  $7.64\pm 0.15$\\
09h-448495$^{f}$  & 0.4080 & $-$19.2 & $1.49\pm 0.11$ & $2.38\pm 0.27$ &
$0.69\pm 0.10$  & U &7.81 & 8.72  &  $8.72\pm 0.15$ \\
09h-392705$^{f}$  & 0.4090 & $-$18.4 & $0.91\pm 0.06$ & $2.94\pm 0.33$ &
$0.85\pm 0.10$  & T &8.16 & 8.51  &  $8.34\pm 0.15$ \\
23h-671455$^{i}$  & 0.4068 & $-$19.0 & $0.41\pm 0.07$ & $1.96\pm 0.13$ &
$0.87\pm 0.05$  & T &8.30 & 8.44   &  $8.37\pm 0.15$\\
23h-683506$^{i}$  & 0.4065 & $-$17.3 & $1.95\pm 0.61$ & $3.05\pm 0.23$ &
$0.77\pm 0.05$  & L &7.90 & 8.65   &  $7.90\pm 0.15$\\
\hline
01h-628521$^{d}$  & 0.6370 & $-$18.7 & $1.23\pm 0.07$ & $2.30\pm 0.22$ &
$0.71\pm 0.02$  & ? &7.86 & 8.70   &  ?\\
01h-584247$^{c}$  & 0.6358 & $-$19.2 & $1.59\pm 0.11$ & $4.03\pm 0.45$ &
$0.89\pm 0.10$  & T &8.11 & 8.51   &  $8.31\pm 0.15$\\
01h-537252$^{c}$  & 0.6372 & $-$19.1 & $1.25\pm 0.09$ & $3.70\pm 0.41$ &
$0.90\pm 0.10$  & T &8.16 & 8.48   &  $8.32\pm 0.16$\\
01h-471561 $^{a}$ & 0.6352 & $-$18.6 & $2.33\pm 0.17$ & $3.70\pm 0.41$ &
$0.84\pm 0.09$  & T &7.97 & 8.60   &  $8.29\pm 0.31$\\
01h-246610$^{b}$  & 0.6390 & $-$17.3 & $12.50\pm 1.45$ & $4.76\pm 0.53$ &
$0.83\pm 0.10$  & L &7.77 & 8.67   &  $7.77\pm 0.15$\\
01h-177719$^{e}$  & 0.6221 & $-$18.4 & $2.86\pm 0.20$ & $5.03\pm 0.56$ &
$0.95\pm 0.08$  & T &8.19 & 8.46   &  $8.33\pm 0.14$\\
23h-441445$^{i}$  & 0.6435 & $-$17.5 & $1.91\pm 0.21$ & $1.67\pm 0.16$ &
$0.53\pm 0.03$  & L &7.53 & 8.85   &  $7.53\pm 0.15$\\
23h-558487$^{i}$  & 0.6433 & $-$18.4 & $2.22\pm 0.27$ & $4.00\pm 0.45$ &
$0.86\pm 0.02$  & T &8.05 & 8.55   &  $8.30\pm 0.15$\\
23h-552534$^{i}$  & 0.6447 & $-$18.3& $2.46\pm 0.25$ & $2.16\pm 0.20$ &
$0.59\pm 0.03$  & L &7.58 & 8.82   &  $7.58\pm 0.15$\\
23h-408246$^{g}$  & 0.6390 & $-$18.6& $6.67\pm 0.47$ & $5.04\pm 0.56$ &
$0.85\pm 0.09$  & L &7.84 & 8.65   &  $7.84\pm 0.15$\\
23h-431392$^{g}$  & 0.6440 & $-$19.2& $1.11\pm 0.08$ & $3.33\pm 0.37$ &
$0.87\pm 0.08$  & T &8.12 & 8.57   &  $8.35\pm 0.23$\\
\hline
23h-385589$^{i}$  & 0.6880 & $-$18.9 & $0.50\pm 0.06$ & $2.25\pm 0.25$ &
$0.88\pm 0.03$  & T &8.29 & 8.44   &  $8.32\pm 0.30$\\
23h-645522$^{i}$  & 0.8903 & $-$19.5 & $0.87\pm 0.08$ & $2.34\pm 0.23$ &
$0.77\pm 0.02$  & T &8.03 & 8.61   &  $8.37\pm 0.15$\\
\hline
\end{tabular}
\end{center}
\end{table*}

\section{The metallicity-luminosity relation at medium redshift}

\begin{figure*}[ht!]
\centerline{\psfig{figure=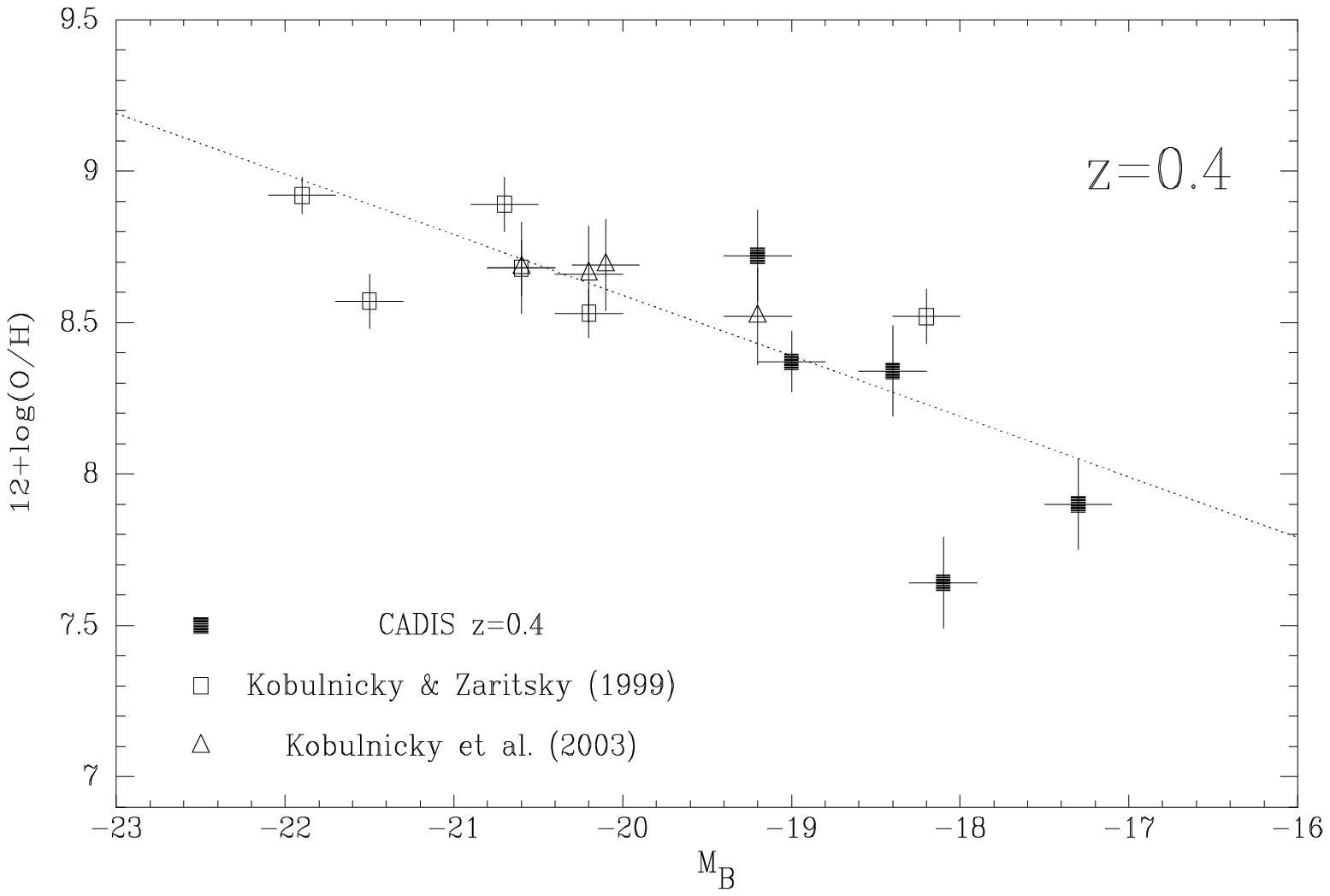,clip=t,width=12cm,angle=0}}\vspace*{1mm}
\centerline{\psfig{figure=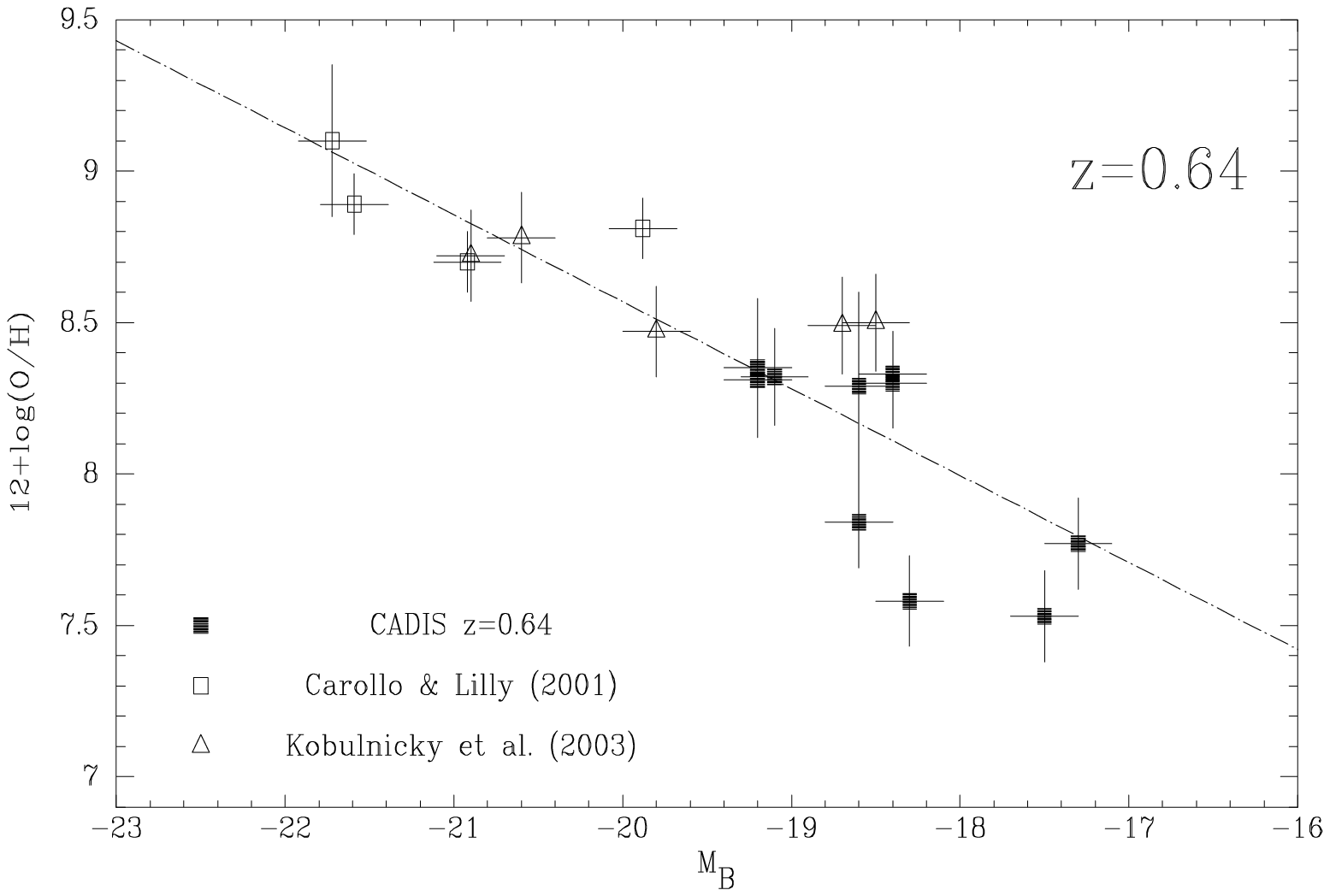,clip=t,width=12cm,angle=0}}
\caption[The oxygen abundance as a
function of absolute magnitude, for $z\sim0.4$  and 
$z\sim0.64$]
{\label{mbfpiab} \footnotesize  The oxygen abundance as a
function of absolute magnitude, for $z\sim0.4$ (upper panel) and
$z\sim0.64$ (lower panel).
The CADIS metallicity measurements are shown as filled squares, open
squares show the abundances found by Kobulnicky \& Zaritsky
(\cite{kobzar}) at $z\approx 0.4$, and by Carollo \& Lilly (\cite{carol}) at $z\approx 0.64$,
for more luminous galaxies.
Open triangles denote measurements from   Kobulnicky et
al. (\cite{kob03}).
The dotted  ($z \sim 0.4$) and dashed-dotted  line ($z \sim 0.64$)   are   linear  fits to the  data points. 
}
\end{figure*}

Oxygen abundances  have been
determined for a total number of 15 CADIS emission line galaxies at medium
redshift with faint
absolute magnitudes ($\rm{M}_{\rm{B}} > -19$). Combining these abundances with published results
for galaxies with brighter absolute magnitudes
we can study the metallicity-luminosity
relation  over a large range of luminosities at  look-back times  of
4.3\,Gyrs ($z \approx0.4$),  and 5.9\,Gyrs ($z \approx 0.64$).
Fig.\,\ref{mbfpiab}  shows the oxygen abundance as a
function of absolute magnitude  for  two redshift bins, at
$z\approx 0.4$, and $z\approx 0.64$, respectively.
Additional to the CADIS measurements, oxygen abundances from literature for galaxies in
the two redshift bins are shown.
It is obvious from the diagrams that only the CADIS emission line
sample provides a meaningful sample of $\rm{M}_{\rm{B}} > -19$ galaxies to study the
faint absolute magnitudes part of the metallicity-luminosity relation
at medium redshift. 

Since our number statistics of oxygen abundances at medium redshift 
 are limited, we applied a linear least squares
fit to the data. 
There may be some curvature in this  correlation. However, it is not
 possible to testify such a curvature with our limited sample.
For emission line galaxies at   $z \approx 0.4$ we find a
metallicity-luminosity relation:
\begin{equation}
\label{melb}
12+\rm{log}(\rm{O/H}) = (4.59\pm0.91) - (0.199\pm0.046) M_{B},
\end{equation}
with an RMS scatter of 0.219.
For emission line galaxies at   $z \approx 0.64$ we find a
metallicity-luminosity relation:
\begin{equation}
\label{melb}
12+\rm{log}(\rm{O/H}) = (2.81\pm0.81) - (0.288\pm0.042) M_{B},
\end{equation}
with an RMS scatter of 0.225.
The dotted ($z \sim 0.4$) and dashed-dotted  line ($z \sim 0.64$) in
Figure \ref{mbfpiab} are  these 
linear  fits  showing that a
metallicity-luminosity relation 
exists at medium redshift
over a larger luminosity range.

\begin{figure*}[ht!]
\centerline{\psfig{figure=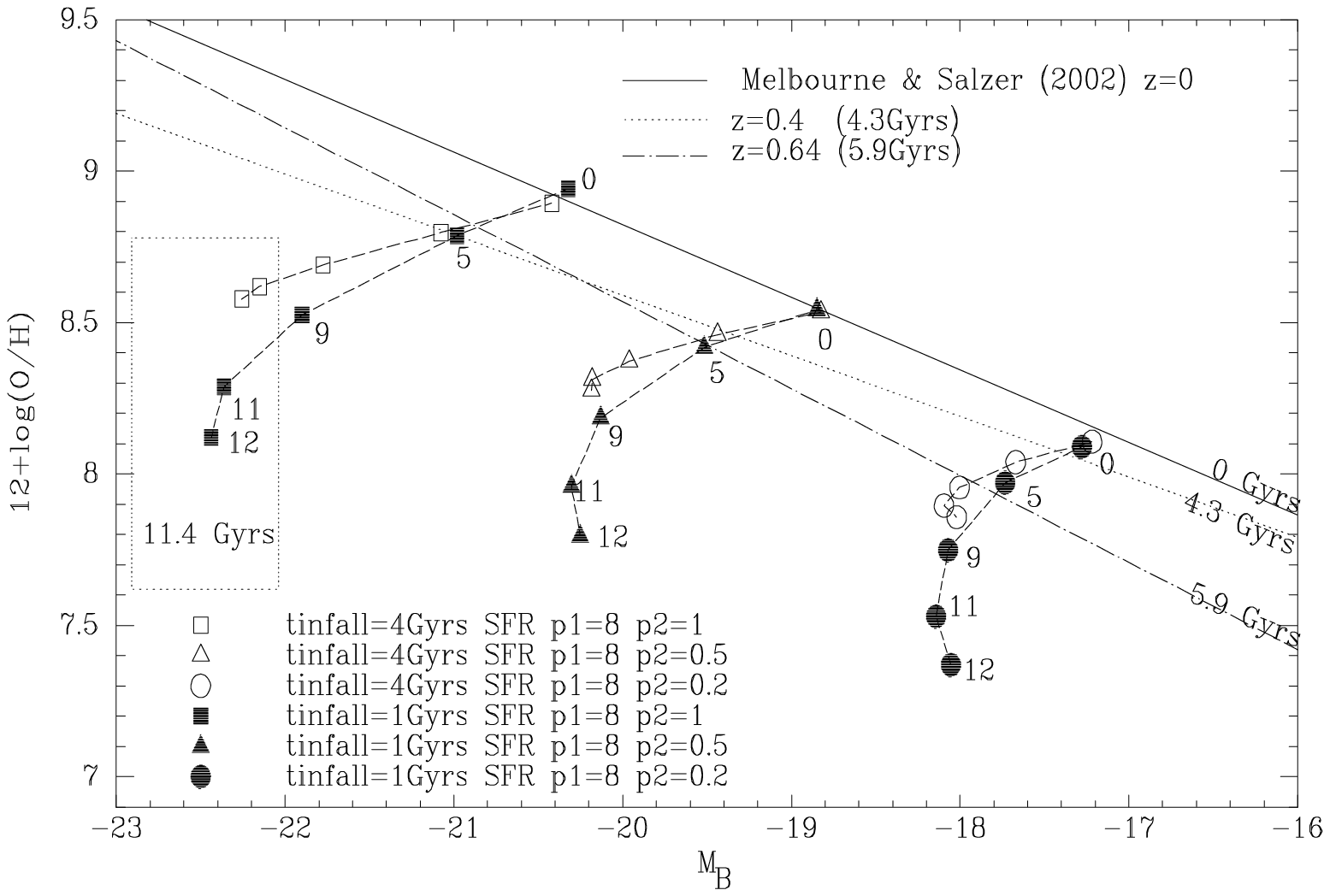,width=12cm,angle=0,clip=true}}
\caption
{\label{verglz} \footnotesize The metallicity-luminosity relation over
  a look-back time of $\sim 12$\,Gyrs.
 The solid line shows the luminosity-metallicity relation found by
Melbourne \& Salzer (\cite{melb}) for a large sample of local
emission-line galaxies (look back time 0\,Gyrs).
Also shown are the metallicity-luminosity relations for galaxies in
at $z\sim0.4$ (look back time 4.3\,Gyrs), and $z\sim0.64$ (look back
  time 5.9\,Gyrs)   from Fig.\,\ref{mbfpiab}.
The location of bright galaxies at $z\sim3.1$ is shown as a box
  encompassing the range of O/H and   $\rm{M_{B}}$  derived for these
  objects (from Fig. 7 in Pettini et
al. (\cite{pettini}), which takes also into account the  metallicities
  derived by  Kobulnicky \& Koo (\cite{kobko}) for $z\sim3.1$ objects). 
The dashed curves and symbols  track  the chemical and luminous evolution of P\'EGASE2 model galaxies with an exponential star formation rate. The galaxies are build by exponentially-decreasing infall of primordial gas (zero metallicity) with infall timescales of 1Gyr and 4Gyrs, respectively. The symbols show galaxies at  a look back time of 0, 1, 5, 9, 11 and 12 Gyrs, assuming that the galaxies begin to assemble at a look back time of 13Gyrs.  
}
\end{figure*}

Fig.\,\ref{verglz} compares the observed  metallicity-luminosity relations at $z=0.4$ and $z=0.64$ (derived in  Fig.\,\ref{mbfpiab}) with the  metallicity-luminosity relation  for $z=0$ emission line galaxies
(Melbourne \& Salzer \cite{melb}) and the oxygen abundances of luminous galaxies
at $z \sim 3.1$ (Kobulnicky \& Koo \cite{kobko}, Pettini et
al. \cite{pettini}). 
Moreover, we compare the observed metallicities and luminosities with P\'egase2 (Fioc \& Rocca-Volmerange \cite{fiocrocca}) models.

\subsection{Comparison to P\'egase2 Models}

P\'egase2 is a galaxy evolution code which allows to study galaxies by evolutionary synthesis.
Here is  a discussion of  the parameters which were chosen by us:

\begin{itemize}

\item
$\varphi(IMF)$: The shape of the stellar initial mass function.
We adopt the Salpeter value, $\alpha=-2.35$, between 0.1 and 120 $M_{\odot}$
(like Baldry et al.  \cite{baldry}, e.g.).

\item
$Y$: The chemical yields from nucleosynthesis.    
We use Woosley \& Weaver \cite{woosw} B-series models for massive
stars.

\item
$M_{tot}$: The total mass of gas available to form the galaxy.
We chose  three representative
masses of $2 \times 10^{10}\,M_{\odot}$,  $6 \times 10^{10}\,M_{\odot}$, and $2 \times 10^{11}\,M_{\odot}$.

\item
$t_{infall}$: The timescale on which the galaxy is assembled.  
We assume that galaxies are built by continuous infall of
primordial gas (zero metallicity) with an infall rate that declines 
exponentially, $t_{infall}$, as implemented in P\'egase2,

\begin{equation}
M(t) = M_{tot} \frac{e^{-t/t_{infall}}}{t_{infall}}.
\end{equation}

\noindent 
We chose two representative gas infall timescales of $t_{infall}=1$\,Gyr and 
$t_{infall}=4$\,Gyrs.

\item
$\psi(SFR)$: The form of the star formation rate. 
We adopt an exponentially decreasing star formation rate, as implemented in P\'egase2:
\begin{equation}
\psi(t) = p2 \frac{e^{-t/p1}}{p1}.
\end{equation}
\noindent

\item
$A_V$: Extinction due to dust.  An inclination-averaged extinction prescription as
implemented in P\'egase2 was included,  but extinction does not play an important role  in the
present comparison (it  changes the model B
magnitudes by only 0.2 mag).

\end{itemize}

We assume that galaxies begin to form about 1 Gyr after the Big Bang
so that local $z=0$ galaxies have  ages of about 13 Gyrs (in a 13.5 
Gyr old universe).
We explored a range of SFR, varying  p1  and p2, and the timescale on which the galaxy is assembled,
$t_{infall}$. 
Assuming  $p1  =  t_{infall}$ a high metallicity is reached after less than 1\,Gyr, and the metallicity do not change much in  the next 12\,Gyrs, not reproducing the observed abundances. 
Models with  $p1 <  t_{infall}$ are unphysical, since SF ceases, although gas is still infalling.
Therefore we explored different values of $p1 >  t_{infall}$. It turned out that, in order to explain the metallicity-luminosity evolution of both luminous galaxies at $z\sim3$  and  galaxies at medium redshift and $z=0$, (representative)  models with values of 
($t_{infall}=4\,\rm{Gyrs}, p1=8$\,Gyrs) or ($t_{infall}=1\,\rm{Gyrs}, p1=8$\,Gyrs), respectively, can reproduce the observed metallicities. The values of $p2$ has been set to $p2=0.2M_{\odot}, p2=0.5M_{\odot}$ and $p2=1M_{\odot}$ according to the total mass of a galaxy of  $2 \times 10^{10}\,M_{\odot}$,  $6 \times 10^{10}\,M_{\odot}$, and $2 \times 10^{11}\,M_{\odot}$, respectively.
Thus, only a SFR history with a long exponential timescale ($\sim 8$\,Gyrs) can reproduce the observations,
but the gas infall has to cease after 1$-$4 Gyrs.

\begin{figure*}[ht!]
\centerline{\psfig{figure=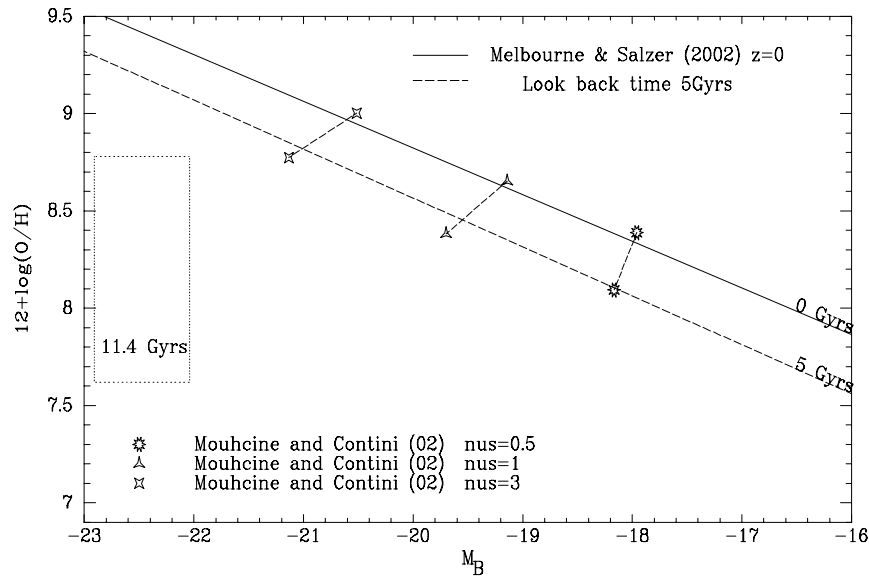,width=12cm,angle=0,clip=true}}
\caption
{\label{sburstmod} \footnotesize 
The metallicity-luminosity relation, similar to Figure \ref{verglz}, but for starbursts models of Mouhcine
 \& Contini (\cite{mouhcont}). Moreover, the dashed line labeled as 5\,Gyrs
 look back time is  the
metallicity-luminosity relation found puting together the metallicities
 of all  emission
line galaxies at $z \sim0.4$ and $z\sim0.64$.}
\end{figure*}

The internal scatter of the metallicity-luminosity relation 
is relatively well understood in the context of  a bursting star
formation mode, shown in  Mouhcine \& Contini
(\cite{mouhcont}).
Therefore, we applied also such a star formation history in our P\'egase2 models:
we  investigated bursting star formation models assuming several
bursts of 50\,Myrs duration with an inter-burst period of 500\,Myrs.
The
star formation rate is assumed to be proportional to the mass of gas to
the power 1.5,
with a star formation efficiency $\nu_{s}$ of 3, 1 and 0.5 Gyr$^{-1}$,
respectively, according to Table 1 in
Mouhcine \& Contini (\cite{mouhcont}) and their equation (3), $ \psi(t)
= \nu_{\rm{s}} \rm{M}_{\rm{gas}}^{1.5}$ for the starbursting phase. The
time for continuous infall of
primordial gas, $t_{infall}$, is assumed to be 4\,Gyrs. Figure \ref{sburstmod} shows,
similar to Figure \ref{verglz},
 the metallicity predicted by P\'egase2 models with these starbursting
SFRs for a look backtime of 5 Gyrs  and for today, compared to the
observed metallicity-luminosity relations.
 Different from Figure \ref{verglz} we show here the
metallicity-luminosity dependence found putting together all the emission
line galaxies at $z \sim0.4$ and $z\sim0.64$ (dashed line in Figure
\ref{sburstmod} labeled as
5\,Gyrs ago). This line corresponds to a
metallicity-luminosity relation:
\begin{equation}
\label{melb}
12+\rm{log}(\rm{O/H}) = (3.55\pm0.60) - (0.251\pm0.031) M_{B},
\end{equation}
with an RMS scatter of 0.227.

Comparison between Figures  \ref{verglz} and \ref{sburstmod} shows that 
 the metallicity-luminosity relation at medium redshift is displaced 
to lower abundances and higher luminosities compared to today. This can be explained 
independent of the
assumption of a starbursting star formation history in the P\'egase2
models or of an exponentially decreasing star formation rate.
The shift we see in the Figures \ref{verglz} and \ref{sburstmod} is
comparable to the internal scatter of the metallicity-luminosity relation
today and at medium redshift. However,  our emission line sample
is not biased compared to the Melbourne and Salzer local sample.
Therefore we think that the shift in our mean relation reflects a true shift,
 in the sense that the metallicity of galaxies  increases
and their luminosity decreases between   $z\sim0.7$ and
$z\sim0$.

\subsection{Discussion}

Galaxies over a large range of  absolute magnitude ($-17 \la
\rm{M}_{\rm{B}} \la -22$) show 
an evolution of the  metallicity-luminosity at $z\sim0.4$ and $z\sim0.64$   compared to $z=0$. 
Comparison with P\'egase2 models shows a possible scenario in which  galaxies at
medium redshift  have faded by $\sim 0.5\,\rm{mag}$  (galaxies with lower luminosities in Fig.\ref{verglz}) up to $\sim 0.9\,\rm{mag}$ (galaxies with higher luminosities in Fig.\ref{verglz})  in $\sim 6$\,Gyrs  due to decreasing levels of
star formation, and their  metallicity has increased by a factor of
$\sim0.3\,\rm{dex}$.

The luminosity evolution of $\Delta
\rm{M_{B}}=0.5-0.9$\,mag between $z=0$ and $z=0.64$  (5.9\,Gyr ago) 
is consistent with the observational results of the COMBO-17 survey (Wolf et al. \cite{wolf}, their Fig.17): They find that 
$\rm{M^{*}_{B}}$ 
of starburst galaxies decreases by $\sim 0.5$\,mag  between $z=0.7$ and $z=0.3$. 

Fig.\,\ref{verglz} and \ref{sburstmod} show an increase of metallicity by a factor of $\sim 2$ for galaxies in the local universe compared to medium redshift.
The metallicity evolution is consistent with models of Somerville \& Primack (\cite{somerv})
and Pei et al. (\cite{pei}), who predict a change in the star-forming
metallicity with redshift of 0.2\,dex, and 0.3\,dex, respectively, over
the last half of the age of the universe (see also Fig.\, 18 in Lilly
et al. \cite{lilly03}).
Our result is also consistent with the finding 
of Lilly et al. (\cite{lilly03}): Based on estimates of the oxygen abundance in a sample of 66 CFRS galaxies at $0.47<z<0.92$  they  argue that \emph{luminous} low metallicity
galaxies at medium redshift, which have normal mass-to-light ratios
and relatively large sizes, will evolve to higher metallicity galaxies
of similar absolute magnitudes
rather than  to fainter low metallicity galaxies.


\subsection*{Acknowledgements}

We are grateful to the anonymous referee for his/her suggestions that
have improved the paper. 
We   thank A. Aguirre and M. Alises from the Calar Alto Observatory for
carring out CADIS observations in service mode, and we thank  Francisco Prada for obtaining TNG observing  time in the
Spanish panel.
We also thank  Eric Bell, Ignacio Ferreras, Alexei Kniazev, Henry Lee and  Simon Lilly  for valuable discussions.


%


\begin{thebibliography}{}

\bibitem[2002]{baldry}   Baldry, I.~K., Glazebrook, K., Baugh, C.~M. et al. 2002, ApJ, 569, 582

\bibitem[1971]{brockle} Brocklehurst M.  1971 MNRAS, 153, 471

\bibitem[2001]{carol} Carollo C. M. \& Lilly S. J. 2001, ApJ, 548, 153 


\bibitem[2002]{cohen} Cohen, J.G. 2002, ApJ, 567, 672

\bibitem[2002]{contini} Contini T., Treyer M. A., Sullivan M. \& Ellis R. S. 2002, MNRAS, 330, 75

\bibitem[2002]{denicolo} Denicolo G., Terlevich R. and Terlevich E. 2002, MNRAS, 330, 69

\bibitem[1999]{fiocrocca}Fioc, M. \& Rocca-Volmerange, B., 1999, astro-ph/9912179

\bibitem[1997]{garnet} Garnett D. R., Shields G. A., Skillman E. D., Sagan S. P. \& Dufour
R. J. 1997, ApJ, 489, 63

\bibitem[1996]{grevesse} Grevesse, N., Noels, A., Sauval, A. J. 1996. Standard Abundances; in
S. S. Holt and G. Sonneborn (Eds.), \emph{Cosmic Abundances:
Proceedings of the 6th annual October
 Astrophysics Conference}, Volume 99 of \emph{A.S.P. Conference
 Series},  p.117.
 San Francisco: Astron. Soc. of the Pacific

\bibitem[2001]{hammer} Hammer F., Gruel N., Thuan T. X. and Infante L. 2001, ApJ, 550, 570 

\bibitem[1992]{hamuy92} Hamuy, M., Walker, A. R., Suntzeff, N. B. et al. 1992, PASP 104, 533

\bibitem[1994]{hamuy94} Hamuy, M, Suntzeff, N. B., Heathcote, S. R. et al. 1994, PASP 106, 566


  
\bibitem[2003]{hippe} Hippelein, H., Maier, C., Meisenheimer, K.  et
 al.
 2003, A\&A, 402, 65

\bibitem[1986]{horne} Horne K. 1986,  PASP, 98, 609

  
\bibitem[1999]{hunter} Hunter D. A. \& Hoffman L. 1999, AJ, 117, 2789


\bibitem[1994]{izo94} Izotov Y. I., Thuan Trinh T. \& Lipovetsky V. A. 1994, ApJ, 435, 647
  
\bibitem[1996]{kauffm} Kauffmann, G. 1996 MNRAS, 281, 475

\bibitem[1998]{kennicut92} Kennicut, R.C.Jr. 1998, ApJ, 498, 541


\bibitem[2004]{kniazev} Kniazev, A.Y., Pustilnik, S.A., Grebel, E.K. et
  al. 2004, 
\emph{Strong Emission Line HII Galaxies in the Sloan Digital Sky Survey.
I. Catalog of DR1 Objects with Oxygen Abundances from Te Measurements}, 
submitted to ApJS
 
\bibitem[2003]{kob03}  Kobulnicky, H. A., Wilmer C. N. A., Weiner,
  B. J. et al.  2003, ApJ, 599, 1006

\bibitem[2000]{kobko}  Kobulnicky, H. A. \& Koo, D. C. 2000, ApJ, 545, 712

\bibitem[1999]{kob99} Kobulnicky, H. A., Kennicutt R. C. Jr. \& Pizagno J. L. 1999, ApJ, 514, 544

\bibitem[1999]{kobzar} Kobulnicky, H. A. \& Zaritsky D. 1999, ApJ, 511, 118

\bibitem[2003]{lilly03} Lilly, S.J, Carollo, C.M. \& Stockton,
  A.N. 2003, ApJ, 597, 730
  
\bibitem[1999]{maclow}   MacLow, M. \& Ferrara, A. 1999, ApJ, 513, 142

\bibitem[1985]{mccall}  McCall L. M., Rybski P. M. \& Shields G. A. 1985, ApJS, 57, 1

\bibitem[1991]{mcgaugh}  McGaugh, Stacy 1991, ApJ, 380, 140

 
\bibitem[2003]{maier03} Maier, C.,  Meisenheimer, K., Thommes, E. et
  al. 2003, A\&A, 402, 79


\bibitem[1998]{meise98} Meisenheimer, K., Beckwith, S., Fockenbrock, H. et al. 1998, in
\emph{The Young Universe: Galaxy Formation and Evolution at Intermediate and 
High Redshift}. Edited by S. D'Odorico, A. Fontana, and E.
Giallongo. ASP Conference Series, Vol. 146,  p.134

\bibitem[2004]{meise03} Meisenheimer, K. et al. 2004, \emph{The Calar Alto Deep Imaging Survey: Concept, Data
Analysis and Calibration}, in preparation

\bibitem[2002]{melb} Melbourne, J. \& Salzer, J. J. 2002, AJ, 123, 2302

\bibitem[2002]{mouhcont} Mouhcine, M. \& Contini, T. 2002, A\&A, 389, 106

\bibitem[1990]{oke} Oke, J. B.   1990, AJ, 99, 1621

  
\bibitem[1989]{osterb} Osterbrock, D.E. 1989, Astrophysics of Gaseous Nebulae and Active
Galactic Nuclei (Mill Valley: University Science Books)

  
\bibitem[1979]{pagel}  Pagel, B. E. J., Edmunds, M. G., Blackwell,
  D. E. et al. 1979, MNRAS, 189, 95

\bibitem[1997]{pagel97} Pagel.,  B. E. J., \emph{Nucleosynthesis and chemical evolution of galaxies}, Cambridge : Cambridge University Press, 1997

\bibitem[1999]{pei} Pei, Y.C., Fall, M., Hauser, M.G. 1999 ApJ, 522, 604
  
\bibitem[2001]{pettini} Pettini, M., Shapley, A. E., Steidel,
  C. C. 2001, AJ, 554, 981

\bibitem[1979]{seaton} Seaton, M. J. MNRAS 1979, 187, 73

\bibitem[1990]{shields} Shields, G. A. ARAA, 1990, 28, 525

\bibitem[1999]{somerv} Somerville, R. S., \& Primack, J. R. 1999,  MNRAS, 310, 1087 

\bibitem[1996]{stasinsk} Stasinska, G. \& Leitherer, C. 1996, ApJS,
  107, 661

\bibitem[1995]{richer} Richer, M. G. \& McCall, M. L. 1995, ApJ, 445, 642

\bibitem[1989]{skilm} Skillman, E. D., Kennicutt, R C. \& Hodge, P.W. 1989, ApJ, 347, 875

\bibitem[1993]{skillm93}  Skillman, E. D. \& Kennicutt, R C. Jr. 1993, ApJ, 411, 655

\bibitem[2003]{wolf} Wolf, C., Meisenheimer, K., Rix, H.-W.  et al 2003, A\&A, 401, 73 

\bibitem[1995]{woosw} Woosley, S. E. \& Weaver, T. A. 1995, ApJS, 101, 181

\bibitem[1958]{whitford}  Whitford, A. E., 1958, AJ,  63, 201

\bibitem[1994]{zarit} Zaritsky, D., Kennicutt, R.C. \& Huchra, J.P. 1994, ApJ, 420, 87

\end{thebibliography}
\end{document}